\documentclass[12pt]{article}
\usepackage{graphicx}
\usepackage{verbatim}
\usepackage{color}
\usepackage{natbib}
\usepackage{setspace}
\usepackage{booktabs}
\usepackage{multirow}
\usepackage{amsthm}

\usepackage{multicol}        

\newcommand{\tr}{^{\prime}}
\def\b#1{\mbox{\boldmath $#1$}}    
\def\cg#1{\mbox{${\cal #1}$}}      
\def\cgl#1{\mbox{\scriptsize {${\cal #1}$}}}

\newcommand{\be}{\beta}

\newcommand{\la}{\lambda}
\newcommand{\La}{\Lambda}

\def\b#1{\mbox{\boldmath $#1$}}

\begin{document}

\title{\vspace*{-1.5cm}Bayesian inference for a class of latent Markov models for
categorical longitudinal data}

\author{Francesco Bartolucci\footnote{Department
of Economics, Finance and Statistics, University of Perugia, IT}
\footnote{email: bart@stat.unipg.it} $\:$ and $\:$ Silvia
Pandolfi$^*$\footnote{email: pandolfi@stat.unipg.it}}

\maketitle\vspace*{-1cm}

\begin{abstract}
\noindent  We propose a Bayesian inference approach for a class of
latent Markov models. These models are widely used for the analysis
of longitudinal categorical data, when the interest is in studying
the evolution of an individual unobservable characteristic. We
consider, in particular, the basic latent Markov, which does not
account for individual covariates, and its version that includes
such covariates in the measurement model. The proposed inferential
approach is based on a system of priors formulated on a
transformation of the initial and transition probabilities of the
latent Markov chain. This system of priors is equivalent to one
based on Dirichlet distributions. In order to draw samples from the
joint posterior distribution of the parameters and the number of
latent states, we implement a reversible jump algorithm which
alternates moves of Metropolis-Hastings type with moves of
split/combine and birth/death types. The proposed approach is
illustrated through two applications based on longitudinal
datasets.\vspace*{0.5cm}


\noindent
\end{abstract}\newpage

\section{Introduction}
The class of latent Markov (LM) models was introduced by
\cite{Wig:55,Wig:73} for the analysis of categorical longitudinal
data. These models are specially tailored to study the evolution of
an individual characteristic which is not directly observable. The
basic LM formulation is similar to that of hidden Markov (HM) models
for time series data \citep{macd:zucc:97}; in fact, a latent Markov
chain, typically of first order, is used to represent the evolution
of the latent characteristic over time. Moreover, the response
variables observed at the different occasions are assumed to be
conditionally independent given this chain (assumption of {\em local
independence}). The basic idea behind this assumption is that the
latent process fully explains the observable behavior of a subject.
Furthermore, the latent state to which a subject belongs at a
certain occasion only depends on the latent state at the previous
occasion. An LM model may also be seen as an extension of the latent
class (LC) model \citep{laza:hen:68, good:74}, in which the
assumption that each subject belongs to the same latent class
throughout the survey is suitable relaxed.

Typical applications of LM models are in studies of the human
behaviors and conditions in health, education, sociology, and
criminology. These models have also been adopted in economics to
study the job market or customer's choice problems. In addition to
Wiggins, among the first authors dealing with LM models, it is also
worth mentioning \cite{vand:delee:86}, \cite{vand:lang:90},
\cite{coll:wug:92}, and \cite{lang:vand:94}. For a complete review
of the state of art of the LM models, see \cite{bart:et:al:10}.

The basic LM model, relying on a homogenous Markov chain, has
several extensions based on parameterizations that allow us to
include hypotheses and constraints of interest. Generally speaking,
these parameterizations may concern the conditional distribution of
the response variables given the latent process ({\em measurement
model}), and/or the distribution of the latent process ({\em latent
model}). An example of an LM model based on constraints on the
measurement model is given by the LM Rasch model
\citep{bart:et:al:08}, which is a generalization of the model
introduced by \cite{rasch:61} allowing each subject to evolve in
his/her ability level. About the latent model, the most interesting
constraints may be expressed on the transition matrix. In
particular, transitions between two given states may be excluded,
and/or certain elements of the transition matrix may be constrained
to be equal \citep[see, among others,][]{bart:06,ver:et:al:99}.
These parametrizations may also be exploited to include individual
covariates in the measurement or in the latent model. The case of
covariates included in the measurement model was dealt with by
\cite{bart:farc:09} among others, whereas the case of covariates
included in the latent model, so that they affect the initial and
the transition probabilities of the Markov chain, was dealt with by
\cite{ver:et:al:99} and \cite{bart:et:al:07b}. In the present paper,
we focus on LM models with constraints and individual covariates
included in the measurement model only.

In the frequentist approach, estimation of the parameters of an LM
model is typically based on the maximum likelihood approach through
the Expectation -Maximization (EM) algorithm
\citep{baum:70,demp:et:al:77}, whose implementation makes use of
suitable recursions. As typically happens for latent variable
models, the likelihood function of an LM model may be multimodal,
and the search of the global maximum may be cumbersome, also due to
the slowness to converge of the EM algorithm. Moreover, this kind of
literature has not still provided a commonly accepted criterion for
formal assessment of the number of the states of the latent chain,
although information criteria are typically used. We refer, in
particular, to the Akaike Information Criterion (AIC), see
\cite{aka:73}, and the Bayesian Information Criterion (BIC), see
\cite{schw:78}. On the other hand, in the Bayesian inference
approach, parameter estimation does not suffer from the problem of
multimodality of the likelihood and model selection is well
principled. Obviously, the condition to apply this type of inference
is that one is able to draw samples from the joint posterior
distribution of the model parameters and the number of latent
states.

In this paper, we propose a Bayesian inference approach for the
basic LM model and its extended versions based on suitable
parametrizations of the conditional response probabilities given the
latent states. These parametrizations may be used to formulate
hypotheses of interest or include individual covariates. The
approach is based on a system of priors that we propose following
the approach for HM models of \cite{capp:et:al:05} and
\cite{spez:10}. Instead of formulating the prior distributions
directly on the initial and transition probabilities of the Markov
chain, we formulate these distributions on unnormalized versions of
these probabilities. In particular, we assume that each unnormalized
initial and transition probability {\em a priori} has an independent
Gamma distribution with suitable hyperparameters. This system of
priors considerably facilitates Bayesian model estimation from the
practical point of view, while being equivalent to a system of
priors based on Dirichlet distributions on the normalized
probabilities.

Under the above system of priors, we estimate the model parameters
and select the number of latent states through a reversible jump
(RJ) algorithm \citep{Green:95}. As is well known, this is a Markov
chain Monte Carlo (MCMC) algorithm which represents an extension of
the Metropolis-Hastings algorithm \citep{metrop:53, hast:70} that
allows us to simulate samples from the posterior distribution when
the
parameter space has varying dimension. 
%
%
Our implementation of the RJ algorithm follows that proposed by
\cite{rich_green:97} for Bayesian estimation of finite mixture
models and that of the RJ algorithm of \cite{robe:ryde:titt:00} for
estimation of HM models. In particular, this implementation is based
on a series of transdimensional moves (i.e., split/combine and
birth/death moves), which allow us to change the number of latent
states. These moves are alternated with moves of MH type to draw
samples from the posterior distribution of the model parameters,
when the same number of latent states is held fixed.


The paper is organized as follows. Section~\ref{sec:basic}
illustrates the basic LM model for univariate and multivariate
categorical longitudinal data, whereas its extensions are discussed
in Section~\ref{sec:ext}. The proposed system of priors is
illustrated in Section~\ref{sec:Bay}. In Section~\ref{sec:mcmc} we
describe the RJ algorithm to draw samples for the posterior
distribution of the model parameters and the number of latent
states. Finally, two applications are illustrated in
Section~\ref{sec:ex}. These applications are based on a dataset
about marijuana consumption and a dataset about female labour
participation. Finally, in Section~\ref{sec:concl} we draw main
conclusions about the proposed approach.

\section{Basic latent Markov model}\label{sec:basic}

We introduce the preliminary concepts about the basic LM model for
categorical longitudinal data, in which the conditional distribution
of each response variable given the corresponding latent variable
and the initial and transition probabilities of the latent process
are unconstrained.

\subsection{Formulation of univariate responses}

In the univariate case, let $\b Y_i =
(Y_i^{(1)},\ldots,Y_i^{(T)}),\, i=1,\ldots,n$, denote a sequence of
$T$ categorical response variables with $l$ levels or categories,
coded from 0 to $l-1$, independently observed over $n$ subjects,
that correspond to repeated measurements on the same subject at
different occasions.

The main assumption underlying the basic LM model is that of {\em
local independence}, i.e. for every subject the response variables
are conditionally independent given a latent process $\b U_i =
(U^{(1)}_i,\ldots,U^{(T)}_i)$. This latent process is assumed to
follow a first-order Markov chain with state space $\{1,
\ldots,k\}$. Then, for all $t>2$, the latent variable $U^{(t)}_i$ is
conditionally independent of $U^{(1)}_i,\ldots,U^{(t-2)}_i$ given
$U^{(t-1)}_i$.

Parameters of the model are the conditional response probabilities
$\phi^{(t)}_{y|u} = p(Y^{(t)}_i = y|U^{(t)}_i = u), \, t = 1,\ldots,
T,\, u = 1,\ldots,k,\, y = 0,\ldots,l - 1$, the initial
probabilities $\pi_u = p(U^{(1)}_i = u),\, u = 1,\ldots,k$, and the
transition probabilities $\pi_{v|u} = p(U^{(t)}_i = v |U^{(t-1)}_i =
u),\, t = 2,\ldots,T,\, u,v = 1,\ldots,k$. Note that the latent
process is time homogeneous, so that the transition probabilities do
not depend on $t$, moreover the initial probabilities are completely
unconstrained. Furthermore, all these probabilities do not depend on
$i$ since, in its basic version, the model does not account for
individual covariates.

The assumptions above imply that the distribution of $\b U_i$ may be
expressed as
$$
p(\b U_i = \b u) =
\pi_{u^{(1)}}\prod_{t>1}\pi_{u^{(t)}|u^{(t-1)}},
$$
where $\b u = (u^{(1)},\ldots,u^{(T)})$. Moreover, the conditional
distribution of $\b Y_i$ given $\b U_i$ may be expressed as
$$
p(\b Y_i = \b y |\b U_i = \b u) = \prod_t
\phi^{(t)}_{y^{(t)}|u^{(t)}},
$$
and, consequently, for the {\em manifest distribution} of $\b Y_i$
we have
\begin{eqnarray}\label{eq:manif}\nonumber
f(\b y)\!\! &=& \!\!p(\b Y_i = \b y) = \sum_u p(\b Y_i = \b y|\b U_i
= \b u)p(\b U_i = \b u) =  \\\!\! & = & \!\! \sum_{u^{(1)}}
\phi^{(1)}_{y^{(1)}|u^{(1)}}\pi_{u^{(1)}}
\sum_{u^{(2)}}\phi^{(2)}_{y^{(2)}|u^{(2)}}
\pi_{u^{(2)}|u^{(1)}}\ldots \sum_{u^{(T)}}
\phi^{(T)}_{y^{(T)}|u^{(T)}}\pi_{u^{(T)}|u^{(T-1)}},
\end{eqnarray}
where $\b y = (y^{(1)},\ldots,y^{(T)})$. In order to efficiently
compute $f(\b y)$ we can use a forward recursion \citep{baum:70},
for obtaining $q^{(t)}(u,\b y) =p(U^{(t)}_i = u, Y^{(1)}_i =
y^{(1)},\ldots,Y^{(t)}_i = y^{(t)})$ for $t = 1,\ldots,T$. The
recursion is as follows: given $q^{(t-1)}(u,\b y)$, $u=1,\ldots,k$,
the t-$th$ iteration consists of computing $$q^{(t)}(v,\b
y)=\phi_{y^{(t)}|v}^{(t)}\sum_u q^{(t-1)}(u,\b y) \pi_{v|u} , \; v
=1,\ldots,k,$$ starting with $q^{(1)}(u,\b
y)=\pi_u\phi_{y^{(1)}|u}^{(1)}$. We then have, $$f(\b y) = \sum_u
q^{(T)}(u,\b y).$$ The above recursion may be efficiently
implemented using the matrix notation, and let $f(\b y) = \b
q^{(T)}(\b y)\tr\b 1$ where $\b 1$ is a column vector of ones of
suitable dimension and $\b q^{(t)}(\b y)$ is a column vector with
elements $q^{(t)}(u,\b y)$. The recursion is then expressed as:
\begin{equation}\label{eq:rec}
\b q^{(t)}(\b y) = \left\{\begin{array}{ll}
\mbox{diag}[\b\phi_{y^{(1)}}^{(1)}]\b\pi, & \quad \mbox{if}\quad t=1, \\
\mbox{diag}[\b\phi_{y^{(t)}}^{(t)}]\b \Pi^{\prime}\b q^{(t-1)}(\b
y), &\quad \mbox{otherwise}.
\end{array}\right.
\end{equation}
with $\b \pi = \{\pi_u,\, u=1,\ldots,k\}$ denoting the initial
probability vector, $\b \phi_y^{(t)} =
\{\phi^{(t)}_{y|u},\,u=1,\ldots,k\}$ denoting the conditional
probability vector and $\b \Pi = \{\pi_{v|u},\, u,v = 1,\ldots,k\}$
denoting the transition probability matrix.

Finally, for an observed sample of $n$ subjects, let $\b y_i$ denote
the observed response vector provided by subject $i$, the model
likelihood may be formulated as $L(\b y|\b \theta) = \prod_i f(\b
y_i)$, where $\b\theta$ is the vector of all model parameters
arranged in a suitable way.

\subsection{Multivariate version}

In the multivariate case, we observe a vector of $r$ response
variables, denoted by $\b Y_i^{(t)}= (Y_{i1}^{(t)},\ldots,
Y_{ir}^{(t)})$, for every subject $i$ and occasion $t$, with
$i=1,\ldots,n$ and $t=1,\ldots,T$. Each response variable has $l_j$
categories, $j=1,\ldots,r$, coded from 0 to $l_j - 1$. Moreover, all
responses provided by subject $i$ are collected in the vector $\b
Y_i$. The assumption of local independence is usually formulated by
also requiring that the elements of each vector $\b Y_i^{(t)}$ are
conditional independent given $U_i^{(t)}$.

The model assumptions imply that
\begin{equation}\label{eq:multi}
p(\b Y_i = \b y |\b U_i = \b u) = \prod_t p(\b Y_i^{(t)} = \b
y^{(t)} | U_i^{(t)} = u^{(t)}),
\end{equation}
where $\b y$ is made of the subvectors $\b y^{(t)} =
(y_1^{(t)},\ldots,y_r^{(t)})$ and $$ p(\b Y_i^{(t)} = \b y^{(t)} |
U_i^{(t)} = u^{(t)}) = \prod_j \phi_{j,y^{(t)}|u^{(t)}}^{(t)},$$
with $\phi_{j,y|u}^{(t)} = p(Y_{ij}^{(t)} = y|U_i^{(t)} = u)$,
$j=1,\ldots,r$, $t = 1,\ldots, T$, $u = 1,\ldots,k$. The manifest
probability $f(\b y)$ has the same expression as in
(\ref{eq:manif}), with $p(\b Y_i = \b y |\b U_i = \b u)$ computed as
in (\ref{eq:multi}), and it may be computed by exploiting the
recursion rule, along similar line as in (\ref{eq:rec}). The
likelihood has the same expression as in the univariate categorical
data.

\section{Constrained and extended versions of the basic model}
\label{sec:ext}

In the basic LM model outlined in the previous section, all the
probabilities are completely unconstrained. There are two
generalizations which may be of interest and commonly arise in
applications. First, we may put restrictions on the parameter space,
in order to give a more parsimonious and easily interpretable model.
Secondly, we may have observed covariates together with the
outcomes. Both generalizations may concern either the distribution
of the response variables (i.e., the {\em measurement model}) or the
distribution of the latent process (i.e., the {\em latent model}).
For a more detailed description see \cite{bart:et:al:10}.

\subsection{Constrained versions}

We discuss here only the constraints on the measurement model in
order to parameterize the conditional response probabilities. In the
univariate case a sensible constraint may be
\begin{equation}\label{eq:const1}
\phi_{y|u}^{(t)} = \phi_{y|u}, \quad t=1,\ldots,T, \; u=1,\ldots,k,
\; y =0,\ldots,l-1.
\end{equation}
This constraint corresponds to the hypothesis that the distribution
of the response variables only depends on the corresponding latent
variable and there is no dependence of this distribution on time.

Other interesting constraints may be expressed by
\begin{equation}\label{eq:const2}
\b\eta_u^{(t)} = \b Z_u^{(t)} \b\beta,
\end{equation}
where $\b\eta_u^{(t)} = \b g(\phi_{0|u}^{(t)},\ldots,
\phi_{l-1|u}^{(t)})$, with $\b g(\cdot)$ being a suitable link
function, $\b Z_u^{(t)}$ being a design matrix and $\b\beta$ being a
vector of parameters.

In the case of binary response variables, we can parameterize the
conditional probabilities through the logit link function
$\eta_u^{(t)} = \log (\phi_{1|u}^{(t)}/\phi_{0|u}^{(t)})$. With
response variables having more than two categories, a natural choice
is that of multinomial logit link function, so that $\eta_u^{(t)}(y)
= \log(\phi_{y|u}^{(t)}/\phi_{0|u}^{(t)}), \; y = 1,\ldots,l-1$.
However when the response variables have an ordinal nature, global
or continuation type logits are more suitable; 
see \cite{bart:06}. For instance, in the case of binary variables,
by assuming that
$$\eta_u^{(t)} = \zeta_u
- \omega^{(t)}, \quad t = 1,\ldots,T, \; u = 1,\ldots,k, $$ we can
formulate a LM version of the Rasch model \citep{rasch:61}, which
finds a natural application in psychological and educational
assessment. In this case, the parameters $\zeta_u$ are interpreted
as ability levels.

%

In the multivariate case, constraints (\ref{eq:const1}) becomes
\begin{equation}
\phi_{j,y|u}^{(t)} = \phi_{j,y|u}, \;\; j = 1,\ldots,r, \; t =
1,\ldots,T, \; u = 1,\ldots,k, \; y = 0,\ldots,l_j-1.
\end{equation}
Moreover, we can use a link function of the type
\begin{equation}\label{eq:const3}
\b \eta_{j,u}^{(t)} = \b Z_{j,u}^{(t)}\b \be,
\end{equation}
in order to parameterize the conditional distribution of each
response variable as in (\ref{eq:const2}), with $\b \eta_{j,u}^{(t)}
= \b g_j(\phi_{j,0|u}^{(t)},\ldots,\phi_{j,l_j-1|u}^{(t)})$.


\subsection{Extended versions based on the inclusion of individual covariates}

As discussed above, the covariates can be included both in the
measurement model and in the latent model. In the former case, the
conditional distribution of the response variables given the latent
states may be parameterized by generalized logits. This
parametrization recalls that used in (\ref{eq:const2}), for the
univariate case, and in (\ref{eq:const3}), for the multivariate
case, to formulate the constraints on the measurement models. Note
that, using this formulation, the assumption of local independence
is relaxed by allowing association between the response variables
observed at the same occasion even conditional on the latent state.

About the model interpretation, when the covariates are included in
the measurement model, the latent process is seen as a way to
account for the unobserved heterogeneity between subjects. The
advantage with respect to a standard random effect or latent class
model with covariates is that we admit that the effect of
unobservable covariates could be non constant over time, but it
could have its own dynamics.

When the covariates influence initial and transition probabilities
of the latent process, we suppose that the response variables
measure and depend on the latent variable (e.g. the quality of
life), which may evolves over time. In such a case, the main
research interest is in modeling the effect of covariates on this
latent variable distribution \citep{bart:et:al:09}.

In this paper, we deal with a model very similar to that proposed by
\cite{bart:farc:09}, that is a multivariate extension of the basic
LM model in which the conditional distribution of the response
variables depends on the individual covariates. This extension is
illustrated in the following.

Let $\b x_i^{(t)}$, denote the vector of individual covariates for
subject $i$ at occasion $t$. Following the formulation of
\cite{bart:farc:09}, we parameterize the conditional distribution of
the response variables given the latent process by a multivariate
marginal link function \citep{bart:et:al:07a}. In particular, let
$\b p_{i,u}^{(t)}$ denote the column vector having elements $p(\b
Y_i^{(t)}=\b y|U_i^{(t)}=u,\b x_i^{(t)})$ for all the possible
configurations $\b y$ of the responses. This probability vector is
parameterized by marginal logits and marginal log-odds ratios which
are collected in the vector $\b\eta_{i,u}^{(t)}$ that may be simply
expressed as
\begin{equation}
\b\eta_{i,u}^{(t)}=\b C\log[\b M\b p_{i,u}^{(t)}],
\end{equation}
where $\b C$ and $\b M$ are appropriate matrices whose construction
is described in \cite{bart:et:al:07a}; see also \cite{col:forc:01}.
Logits and log-odds ratios may be of local, global, or continuation
type; the choice is driven by the nature of the response variables,
essentially ordinal or non-ordinal.

To relate the above marginal effects to the covariates, we assume
that \begin{equation}\label{eq:assump} \b\eta_{1,i,u}^{(t)} =
\b\xi_u + \b X_i^{(t)}\b\beta, \qquad \b\eta_{2,i,u}^{(t)} =
\b\gamma,
\end{equation}
where $\b\eta_{1,i,u}^{(t)}$ is the subvector of
$\b\eta_{i,u}^{(t)}$ containing the logits and
$\b\eta_{2,i,u}^{(t)}$ is the subvector containing the log-odds
ratios. Moreover, $\b X_i^{(t)}$ is a suitable design matrix defined
on the basis of $\b x_{i}^{(t)}$, whereas $\b\beta$ and $\b\gamma$
are vectors of parameters. 
Note that $\b\xi_u$, $u=1,\ldots,k$, may be seen as
support points, corresponding to each latent states, for individual
random effects which are time-varying.

As discussed above, the resulting model allows for unobserved
heterogeneity beyond individual covariates; moreover, the effect of
the first is admitted to be time-varying. This extension is of
interest when we want to investigate on the direct effect of the
covariates on the response variables.

%
\section{Bayesian setting}\label{sec:Bay}
The basic LM model and its extended versions are considered here in
the Bayesian setting; at this aim, we introduce the system of priors
elicited for the dimension and the unknown model parameters.

\subsection{Basic latent Markov model}\label{sec:bay_basic}
In specifying the prior distributions for the initial and transition
probabilities we follow the approach of \cite{capp:et:al:05} and
\cite{spez:10}, who exploit a transformation (based on unnormalized
probabilities) which facilitates the estimation. In particular, we
let $\pi_u=\lambda_u/\sum_v\lambda_v$ where $\lambda_u$, are assumed
{\em a-priori} independent, with distribution $Ga(\delta_u,1)$ for
$u=1,\ldots,k$; similarly, $\pi_{v|u} = \lambda_{uv}/\sum_w
\lambda_{uw}$ where $\lambda_{uv}$, are assumed {\em a-priori}
independent, with distribution $Ga(\delta_{uv},1)$ for
$u,v=1,\ldots,k$. The $\lambda_{uv}$ are not identified, but this
transformation facilitates the MCMC moves, since it relaxes the
constraints on the initial and transition probabilities \citep[see
also][]{capp:et:al:03}.
Typically, the hyperparameters are chosen as $\delta_fu=1$,
$u=1,\ldots,k$, and $\delta_{uv} = k\cdot I(u=v)+0.6\cdot I(u \ne
v)$, $u,v=1,\ldots,k$, where $I(A)$ is the indicator function. With
the latter choice, as usual in LM models, the probability of
persistence is greater than the probability of transition. This
system of priors results equivalent to a system based on Dirichlet
distributions. In the following, we denote by $\b\lambda$ and
$\b\Lambda$ the vector and the matrix with elements $\lambda_u$ and
$\lambda_{uv}$, respectively.

We also consider the same reparametrization for the conditional
response probabilities, through the vectors $\b\psi_y^{(t)}$ with
elements $\psi_{yu}^{(t)}$, as
$\phi^{(t)}_{y|u}=\psi_{yu}^{(t)}/\sum_h\psi_{hu}^{(t)}$. We assume
an independent Gamma prior distribution for $\psi_{yu}^{(t)} \sim
Ga(\delta^{(t)}_{yu},1)$, choosing $\delta^{(t)}_{yu}=1$ for $y =
0,\ldots,l-1$, $u=1,\ldots,k$, $t = 1,\ldots,T$.

Finally, for the parameter $k$ we define a discrete Uniform prior
distribution between 1 and $k_{\max}$, where $k_{\max}$ is the
maximum number of states we admit a priori. Usually $k_{\max}$ is
greater than the most complex model that could be visited by the
algorithm; we choose $k_{\max}=10$.

The above setting can be easily extended to the case of multivariate
categorical data.

\subsection{Constrained and extended versions}\label{sec:bay_ext}

In the constrained versions expressed by (\ref{eq:const2}) and
(\ref{eq:const3}), once defined a system of priors for the initial
and transition probabilities and for the number of latent states
$k$, as in Section~\ref{sec:bay_basic}, it only remains to choose a
prior distribution for the generic vector of parameters $\b \be$. It
is natural to assume $\b \be \sim N(\b 0,\sigma^2_\be \b I)$, where
$\b 0$ and $\b I$ are respectively a vector of zeros and an identity
matrix of suitable dimension. The choice of $\sigma^2_\be$ depends
on the constraints adopted and on the context of application.
Typically, $5\leq\sigma^2_{\beta}\leq 10$.

About the model based on assumption (\ref{eq:assump}), which allows
for the inclusion of individual covariates, we assume that the
vectors $\b\xi_u$ are {\em a-priori} independent with distribution
$N(\b 0,\sigma^2_\xi \b I)$. Similarly, we assume that $\b\beta\sim
N(\b 0,\sigma^2_{\beta}\b I)$ and $\b\gamma\sim N(\b
0,\sigma^2_{\gamma}\b I)$. Also in this case we assume
$5\leq\sigma^2_\xi = \sigma^2_{\beta} = \sigma^2_{\gamma}\leq 10$.
Again, concerning the initial and transition probabilities and the
number $k$ of latent states, are still valid the prior assumptions
defined in Section~\ref{sec:bay_basic}.

\section{Reversible jump algorithm}
\label{sec:mcmc} In this paper, we propose a framework for Bayesian
inference on LM models, implementing a RJ algorithm which draws
samples from the posterior distribution of the parameters and
simultaneously from that of the number of latent states. The
proposed framework has many points in common with those developed
for HM models \citep{robe:ryde:titt:00,spez:10} about the
specification of the priors and/or the structure of the estimation
algorithm. 

In particular, the proposed algorithm is based on two different
types of move. The moves of the first type are aimed at updating the
parameters of the current model given the number of states; those of
the second type also allow us to update the number of states. In
more detail, the algorithm performs the following steps:
\begin{description}
\item[Step 1:] Metropolis-Hastings (MH)
move in order to draw, given the current $k$, the parameters from
their posterior distributions.
\item[Step 2:] split/combine move (each proposed with probability
0.5). The split proposal consists of choosing a state at random and
splitting it into two new ones. The corresponding parameters are
split using auxiliary variables. In the combine move a pair of
states is picked at random and merged into a new one, so as to
recover the values of the auxiliary variables of the split move.
\item[Step 3:] birth/death move (each proposed with probability 0.5).
The birth move is accomplished by generating a new state and drawing
the new parameters from their respective priors. In the death move a
state is selected at random and then deleted along with the
corresponding parameters.
\end{description}

This structure closely recalls the one of the RJ algorithm for
mixture models proposed by \cite{rich_green:97}, although the birth
and death moves are not limited to the empty components
\citep[see][Ch. 13]{spez:10,capp:et:al:05}. Moreover, in the first
type of move, the simulation from the posterior density is
accomplished through the MH step instead of implementing a Gibbs
sampler. Furthermore, our implementation does not simulate the
latent process since it directly exploits the manifest (or marginal)
distribution, which is computed by a suitable recursion
\citep{baum:70}. We decide to simulate from the posterior
distribution of the parameters using random walk MH moves without
completion because the resulting algorithm is easier to implement
within the RJ framework. Moreover, as pointed out in
\cite{cel:et:al:00}, the Gibbs sampler is not always appropriate for
sampling from a multimodal distribution, because it is not always
able to explore the posterior surface and to escape local mode
\citep[see also][]{jasra:et:al:05}. Finally, instead of passing
through each step deterministically, we choose to randomly select,
at each iteration, among split/combine and birth/death step with
probability equal to 0.5. The MH step is always performed.

A well-known problem occurring in Bayesian mixture modeling, is the
label switching problem that can be seen as the non-identifiability
of the component due to the invariance of the posterior distribution
to the permutations in the parameters labeling. Several solutions
have been proposed in the literature; in particular, we cite:
artificial identifiability constraints
\citep{dieb:robe:94,rich_green:97} and the related random
permutation sampling \citep{fruh:01}, the relabeling algorithm
\citep{cel:98,steph:00b} and the label invariant loss function
methods \citep{cel:et:al:00,hurn:et:al:03}. For a general review see
\cite{jasra:et:al:05} and \cite{spez:09}. Since this issue is of
great complexity, we decide to use relabeling techniques
retrospectively, by post-processing the RJ output as in
\cite{marin:et:al:05}. In particular, the label switching is managed
by sorting the MCMC sample of the draw obtained at the end of the
iterations on the basis of the permutation of the states which
minimizes the distance from the posterior mode. More details are
given in Section~\ref{sec:algor}.

\subsection{The algorithm}\label{sec:algor}

We now describe in more detail the three steps of the RJ algorithm
which allow for the estimates of the model parameters and the
unknown number of states $k$. Concerning the constrained and the
extended versions, we only illustrate these steps for the
multivariate LM model with covariates formulated on the basis of
assumption (\ref{eq:assump}); the steps for LM versions based on
different assumptions may be easily derived.

\subsubsection*{Step 1-  Metropolis Hastings move}

In the first step of the algorithm, with fixed $k$, the model
parameters are drawn from their posterior distribution on the basis
of separate multiplicative random proposals.
About the unnormalized initial, transition and conditional
probabilities, we consider a logarithmic transformation of the
positive quantities $\lambda_u$ $\lambda_{uv}$, and
$\psi_{yu}^{(t)}$, in order to mapped them onto the real line. The
proposed moves are:
\begin{enumerate}
\item $\log \lambda_u^* = \log \lambda_u + \epsilon_u,\;$ with $\;\epsilon_{u} \sim N(0,\tau_\la)\;$ for $\;u =
1,\ldots,k$.
\item $\log \la_{uv}^* = \log \la_{uv} + \epsilon_{uv},\;$ with $\;\epsilon_{uv}\sim N(0,\tau_\La)\;$ for $\;u,v =
1,\ldots,k$.
\item \begin{enumerate}
\item For the basic LM model,\begin{enumerate}\item[-] $\log \psi_{yu}^{(t)*}  =  \log \psi_{yu}^{(t)} +
 \epsilon_{yu}^{(t)},\;$ with $\;\epsilon_{y,u}^{(t)} \sim N(0,\tau_\psi)\;$ for $y=0,\ldots,l-1,\; u=1,\ldots,k,\\
 t=1,\ldots,T$.\end{enumerate}
\item For the extended model with covariates, \begin{enumerate}
\item[-] $\xi_{hu}^* = \xi_{hu} +
\epsilon_{hu},\;$ with $\;\epsilon_{h,u} \sim N(0,\tau_\xi)\;$ for
$\;h = 1,\ldots,n_\xi,$ $u =1,\ldots,k$.
\item[-] $\beta_i^*  =  \beta_i + \epsilon_i,\;$ with $\;\epsilon_i \sim
N(0,\tau_\be)\;$ for $\;i=1,\ldots,n_\be$.
\item[-] $\gamma_j^* =
\gamma_j + \epsilon_j,\;$ with $\;\epsilon_j \sim N(0,\tau_\gamma)$,
for $\;j = 1,\ldots,n_\gamma$.
\end{enumerate}
\end{enumerate}
\end{enumerate}
Note that $n_{\xi}$, $n_{\be}$ and $n_{\gamma}$ are the dimension of
the vectors $\b\xi_u$, $\b\be$ and   $\b\gamma$, i.e. respectively,
the number of marginal logits, the number of parameters, and the
number of log-odds ratios. The acceptance probabilities of the
proposed values, for both versions, include the Jacobian that arises
because we work with a log-scale transformation. This is given by
$\prod_{u,v}\la_{uv}^*/\la_{uv}$,
$\prod_{u}\lambda_{u}^*/\lambda_{u}$ and
$\prod_{y,u,t}\psi_{yu}^{(t)*}/\psi_{yu}^{(t)}$, respectively.


\subsubsection*{Step 2 - split/combine move}


Suppose that the current state of the chain is $(k,\b\theta_k)$; in
this step we choose between split and combine move with probability
0.5. Obviously, when $k=1$, we always propose a split move, while
when $k=k_{\max}$ we propose a combine move.

In the split move a state $u_0$ is randomly selected and split it
into two new ones, $u_1$ and $u_2$. The corresponding parameters are
split as follows:
\begin{enumerate}
\item Split $\lambda_{u_0}$ as \vspace{-5mm}
$$\lambda_{u_1} = \lambda_{u_0}\,\rho,\quad \lambda_{u_2} =
\lambda_{u_0} (1-\rho)\quad \hbox{with}\;\; \rho \sim U(0,1).$$
\item \vspace{-5mm} Split column $u_0$ of $\b\La$ as \vspace{-5mm}
$$\la_{u \,u_1} =
\la_{u\,u_0}\,\rho_u, \quad \la_{u\,u_2} =
\la_{u\,u_0}(1-\rho_u)\quad \hbox{with} \;\; \rho_u \sim
U(0,1)\;\hbox{for}\;\; u \ne u_0.$$
\item \vspace{-5mm} Split row $u_0$ of $\b \La$ as \vspace{-5mm}
$$\la_{u_1\,v} = \la_{u_0\,v}\,\vartheta_v, \quad \la_{u_2\,v} = \la_{u_0\,v}/\vartheta_v \quad \hbox{with} \;\; \vartheta_v \sim
Ga(a_\la,b_\la)\; \hbox{for} \;\; v \ne u_0.$$
\item \vspace{-5mm} Split $\la_{u_0\,u_0}$ as \vspace{-5mm}
\begin{eqnarray*}
\la_{u_1\,u_1} & = & \la_{u_0\,u_0}\rho_{u_0}\vartheta_{u_1}, \quad
\la_{u_1\,u_2} = \la_{u_0\,u_0}(1-\rho_{u_0})\vartheta_{u_2},\\
\la_{u_2\,u_1}& = & \la_{u_0\,u_0}\rho_{u_0}/\vartheta_{u_1},\quad
\la_{u_2\,u_2} = \la_{u_0\,u_0}(1-\rho_{u_0})/\vartheta_{u_2},
\vspace{-5mm} \end{eqnarray*}
with $\rho_{u_0} \sim U(0,1)$ and
$\vartheta_{u_1}, \vartheta_{u_2} \sim Ga(a_\vartheta,b_\vartheta)$.
\item
\begin{enumerate}
\item For the basic LM model, split $\psi_{y\,u_0}^{(t)}$ as \vspace{-5mm}
$$\psi_{y\,u_1}^{(t)} = \psi_{y\,u_0}^{(t)}\vartheta_y^{(t)}, \quad \psi_{y\,u_2}^{(t)} =
\psi_{y\,u_0}^{(t)}/\vartheta_y^{(t)},$$ with $\vartheta_y^{(t)}
\sim Ga(a_\psi,b_\psi)$ for $y =0,\ldots,l-1, \; t=1,\ldots,T$.
\item For the extended model with covariates, perturbate $\xi_{h\,u_0}$
as \vspace{-5mm}
$$\xi_{h\,u_1} = \xi_{h\,u_0} - \varepsilon_h, \quad \xi_{h\,u_2} =
\xi_{h\,u_0} +\varepsilon_h, \quad \hbox{with} \;\; \varepsilon_h
\sim N(0,\tau\tr_\xi)\; \hbox{for} \;\; h = 1,\ldots, n_{\xi}.$$
\end{enumerate}
\end{enumerate}
\vspace{-5mm} It is worth noting that the densities of the proposals
are identical, as $\rho$ and $1-\rho$ have the same distribution and
likewise for $\vartheta$ and $1/\vartheta$ and $\varepsilon$ and
$-\varepsilon$ respectively (here subscripts on these variables are
omitted), so the symmetry constraints are satisfied.

In the reverse combine move two distinct states, $u_1$ and $u_2$,
are picked at random and merged into a single state $u_0$, so as to
preserve reversibility:
\begin{enumerate}
\item  $\lambda_{u_0} = \lambda_{u_1} + \lambda_{u_2}$.
\item $\la_{u\,u_0} = \la_{u\,u_1}+\la_{u\,u_2} \;$  for  $\;\; u \ne
u_0$.
\item $\la_{u_0\,v} = (\la_{u_1\,v}\,\la_{u_2\,v})^{1/2}\;$ for $ \;\; v \ne
u_0$.
\item $\la_{u_0\,u_0} =
(\la_{u_1\,u_1}\la_{u_2\,u_1})^{1/2}+(\la_{u_1\,u_2}\la_{u_2\,u_2})^{1/2}$.
\item \begin{enumerate}\item For the basic LM model $\psi_{y\,u_0}^{(t)} = (\psi_{y\,u_1}^{(t)}
\psi_{y\,u_2}^{(t)})^{1/2}\;$ for $y = 0,\ldots,l-1$.
\item For the extended model with covariates $\xi_{h\,u_0} = (\xi_{h\,u_1} +
\xi_{h\,u_2})/2$ for $h=1,\ldots,n_{\xi}$.
\end{enumerate}
\end{enumerate}
Note that the split/combine move does not influence the parameters
$\b \beta$ and $\b\gamma$ as they are not affected by the number of
states.

The split move is accepted with probability $\min\{1,A\}$ whereas
the combine move is accepted with probability $\min\{1,A^{-1}\}$. In
the basic LM model, $A$ can be computed as
\begin{eqnarray}\label{eq:A_split}
A = &&\frac{L(\b y|\b\theta_{k+1})p(\b\theta_{k+1})p(k+1)}{L(\b y|\b\theta_k)p(\b\theta_k)p(k)}\times\frac{(k+1)!}{k!}\times\frac{P_c(k+1)/[(k+1)k/2]}{P_s(k)/k}\nonumber\\
&&\times\frac{|J|}{2\,
p(\vartheta_{u_1})p(\vartheta_{u_2})\prod_{v\ne
u_0}p(\vartheta_{v}) \prod_{y}\prod_t p(\vartheta_{y}^{(t)})} \nonumber \\
&&=\frac{L(\b y|\b\theta_{k+1})p(\b\theta_{k+1})}{L(\b
y|\b\theta_k)p(\b\theta_k)}\times\frac{P_c(k+1)}{P_s(k)}\times\frac{|J|}{p(\vartheta_{u_1})p(\vartheta_{u_2})\prod_{v\ne
u_0}p(\vartheta_{v}) \prod_{y} \prod_t p(\vartheta_{y}^{(t)})}.\nonumber\\
\end{eqnarray}
When we deal with the model assumption (\ref{eq:assump}), the above
formula becomes
\begin{eqnarray}\label{eq:A_split2}
A=&&\frac{L(\b y|\b\theta_{k+1})p(\b\theta_{k+1})p(k+1)}{L(\b y|\b\theta_k)p(\b\theta_k)p(k)}\times\frac{(k+1)!}{k!}\times\frac{P_c(k+1)/[(k+1)k/2]}{P_s(k)/k}\nonumber\\
&&\times\frac{|J|}{2\,
p(\vartheta_{u_1})p(\vartheta_{u_2})\prod_{v\ne
u_0}p(\vartheta_{v}) \prod_h p(\varepsilon_h)} \nonumber \\
&&=\frac{L(\b y|\b\theta_{k+1})p(\b\theta_{k+1})}{L(\b
y|\b\theta_k)p(\b\theta_k)}\times\frac{P_c(k+1)}{P_s(k)}\times\frac{|J|}{p(\vartheta_{u_1})p(\vartheta_{u_2})\prod_{v\ne
u_0}p(\vartheta_{v}) \prod_h p(\varepsilon_h)}.\nonumber\\
\end{eqnarray}

In both the equation, $L(\b y|\b\theta_{k})$ represents the
likelihood computed via the forward algorithm, while $p(\b\theta_k)$
is the prior distribution of all model parameters. Moreover,
$P_s(k)/k$ and $P_c(k+1)/[(k+1)k/2]$ are respectively the
probabilities to split a specific component out of $k$ available
ones, and to combine one of $(k+1)k/2$ possible pairs of components.
We also note that $p(k+1)/p(k)$ cancels out and that the Uniform
variables involved have densities equal to unity.
%
The factorials and the coefficient 2 arise from combinatorial
reasoning related to label switching. $|J|$ is the Jacobian of the
transformation from $\b\theta_k$ to $\b\theta_{k+1}$, which is the
product of five determinants $|J_1|  = \lambda_{u_0}$, $\;|J_2| =
\prod_{u \ne u_0} \lambda_{u\,u_0}$, $\;|J_3| = 2^{k-1} \prod_{v \ne
u_0} \la_{u_0\,v}/\vartheta_v$, $\;|J_4| = 4 \la_{u_0\,u_0}^3
\rho_{u_0}(1-\rho_{u_0})/\vartheta_{u_1}\vartheta_{u_2}$ and
$\;|J_{5a}| = \prod_y \prod_t
2\psi_{y\,u_0}^{(t)}/\vartheta_y^{(t)}$ or $\;|J_{5b}| =
2^{n_{\xi}}$.

\subsubsection*{Step 3 - birth/death move}

This step is performed with probability 0.5, along similar lines as
split/combine move.

The birth move is accomplished by generating a new state, denoted by
$u_0$, drawing
the new parameters from their respective priors. 
The remaining parameters are simply copied to the proposed new state
$\b \theta_{k+1}$. In the death move a state $u_0$ is selected at
random and then deleted along with the corresponding parameters.

The acceptance probability of the birth move is $\min\{1,A\}$, where
$A$ may be expressed by the following formulas
\begin{eqnarray}
A=&&\frac{L(\b y|\b\theta_{k+1})p(\b\theta_{k+1})p(k+1)}{L(\b y|\b\theta_k)p(\b\theta_k)p(k)}\times\frac{(k+1)!}{k!}\times\frac{P_d(k+1)/(k+1)}{P_b(k)}\nonumber\\
&&\times\frac{|J|}{p(\la_{u_0})p(\la_{u_0\,u_0})\prod_{u \ne
u_0}p(\la_{u\,u_0})\prod_{v \ne
u_0}p(\la_{u_0\,v}) \prod_y \prod_t \psi_{y\,u_0}^{(t)} },\nonumber\\
\end{eqnarray}
\begin{eqnarray}
A=&&\frac{L(\b y|\b\theta_{k+1})p(\b\theta_{k+1})p(k+1)}{L(\b y|\b\theta_k)p(\b\theta_k)p(k)}\times\frac{(k+1)!}{k!}\times\frac{P_d(k+1)/(k+1)}{P_b(k)}\nonumber\\
&&\times\frac{|J|}{p(\la_{u_0})p(\la_{u_0\,u_0})\prod_{u \ne
u_0}p(\la_{u\,u_0})\prod_{v \ne
u_0}p(\la_{u_0\,v}) \prod_h p(\xi_{h\,u_0}) },\nonumber\\
\end{eqnarray}
that can be applied to the basic LM model and to the model with
covariates, respectively.

The death move is accepted with probability $\min\{1,A^{-1}\}$.
Since the proposal densities are equal to the priors of the
corresponding parameters, and because the components in $\b\theta_k$
remain the same in $\b\theta_{k+1}$, many terms cancel out in the
expression above. Note that $|J|=1$.

\subsubsection*{Post-processing method}

At the end of the iterations of the algorithm, we select the model
with the highest posterior probability of the number of state $k$,
i.e. the model that has been visited most often by the RJ algorithm,
after discarding the burn-in period. After that we collect the MCMC
sample of the draws obtained when the best model was visited. Then
it is possible to compute the ergodic averages of those parameters,
as $\b\be$ and $\b\gamma$, that are not affected by the number of
states. Concerning the remaining model parameter estimates, and in
order to tackle the label switching problem, we need to apply the
post-processing method, as in \cite{marin:et:al:05}. In particular,
for a sample $\b\theta^{(1)},\ldots,\b\theta^{(N)}$ drawn from the
posterior distribution of the parameters of a model with $k$ latent
states, the post-processing method is based on the following steps:
\begin{enumerate}
\item compute the posterior mode $\hat{\b\theta}$ as:
$$\hat{\b\theta}=
{\rm argmax}_{i=1,\ldots,N}\: L(\b y|\b
\theta^{(i)})p(\b\theta^{(i)}).$$
\item Let $h(\b\theta^{(i)})$ denote the vector $\b\theta^{(i)}$ with elements
permuted according to certain permutation of the latent states and
let $\cg H$ denote the space of all possible
permutations.\vspace{0.25cm}
\item For $i=1,\ldots,N$, substitute $\b\theta^{(i)}$ with the
corresponding permutation which minimizes the distance from
$\hat{\b\theta}$, i.e.,
\[
{\rm argmin}_{h\in\cgl H} \|h(\b\theta^{(i)})-\hat{\b\theta}\|.
\]
\end{enumerate}

\section{Empirical illustrations}\label{sec:ex}

To illustrate the Bayesian inference for the class of LM models
proposed in this paper, we describe the analysis of two real
datasets. The first one concerns the use of marijuana among young
people and it is analyzed using the basic LM model and a model with
constraints on the conditional response probabilities. The second
one is a dataset extracted from the database derived from the Panel
Study of Income Dynamics, and is about fertility and female
participation to the labor market. In this case, to fit the data, we
use the more complex LM model based on assumption (\ref{eq:assump}),
which allows for the presence of covariates.

\subsection{Marijuana consumption dataset}\label{sub:drug}

The marijuana consumption dataset has been taken from five annual
waves (1976-1980) of the National Youth Survey \citep{ell:et:al:89}
and is based on $n=237$ respondents who were aged 13 years in 1976.
The use of marijuana is measured through $T=5$ ordinal variables,
one for each annual wave, with $l=3$ levels coded as 0 for never in
the past year, 1 for no more than once a month in the past year and
2 for more than once a month in the past year. We want to explore
whether there is an increase of marijuana use with age.

As illustrated by \cite{ver:hag:04}, a variety of models may be used
for the analysis of this dataset but a LM approach is desirable for
its flexibility and easy interpretation \citep[see][]{bart:06}.

In our implementation, we used the system of priors outlined in
Section~\ref{sec:bay_basic} with  $\delta_u=1$, $\delta_{uv} =
k\cdot I(u=v)+0.6\cdot I(u \ne v)$, and $\delta^{(t)}_{yu}=1$, for
the unnormalized initial and transition probabilities and for the
conditional response probabilities, respectively; $k_{\max}$ was set
equal to 10. Moreover, in the MH step, the parameters were updated
for fixed $k$ through an increment random walk proposal on each
$\log \lambda_u$, $\log \la_{u,v}$, and $\log \psi_{yu}^{(t)}$, with
$\tau_{\lambda}=0.5$, $\tau_\Lambda = 0.1$ and $\tau_\psi=0.2$. The
sampler parameters were tuned so as to achieve acceptance rates in
the range $0.1-0.25$ for all values $k \le 10$. In the split move we
used $a_\la = a_\vartheta = a_\psi = 1$ and $b_\la = b_\vartheta =
b_\psi = 1$ as parameters of the Gamma distributions. We also
decided to simplify the model using constraint (\ref{eq:const1}), in
which the conditional probabilities were assumed to be time
homogeneous, i.e. $\phi_{y|u}^{(t)} = \phi_{y|u}$ and analogously,
$\psi_{yu}^{(t)} = \psi_{yu}$. Finally, the algorithm ran for
$1,000,000$ iterations with a burn-in of $200,000$ iterations. The
starting values were randomly chosen. The acceptance rates are
illustrated in Table \ref{tab:acc}. Concerning the transdimensional
moves, we note that the acceptance rates are a bit lower than
desired, but if we consider that split/combine or birth/death moves
only involve a change of model dimension, and all the other
parameters are updated in each sweep, they are not too low
\citep[see][for comparable results]{robe:ryde:titt:00}.
\begin{table}[h!]\centering\vspace*{0.25cm}
{\small
\begin{tabular}{lrrr}
\toprule
   &Performed &Accepted & \% Accepted\\
\midrule
MH with fixed k & 1,000,000 &  \\
{\em \quad Initial probabilities}& &      209,312 & 20.93 \\
{\em \quad Transition probabilities} & &  127,815 & 12.78\\
{\em \quad Conditional probabilities} & & 133,458 & 13.35\\
Birth  & 250,268 & 2,129 & 0.85 \\
Death  & 250,229 & 2,117 & 0.85 \\
Split &  250,107 & 770 & 0.31 \\
Combine &249,396 & 788& 0.32\\
\bottomrule
\end{tabular}}
\caption{\em Acceptance rates for MH move,  split/combine and
birth/death move under the basic LM model} \label{tab:acc}
\vspace*{0.25cm}
\end{table}%
The estimated posterior probabilities are 0.689, 0.277, 0.031, 0.002
for $k=3,4,5,6$ and below 0.001 for smaller and larger values of
$k$. Hence, the most probable model is that with three latent state,
the same as selected by \cite{bart:06} using BIC.

In order to face the label switching problem, at the end of all
iterations, we post-processed the output as illustrated in Section
\ref{sec:algor}. Moreover, to have a clearer interpretation of the
results, the MCMC draws were sorted on the basis of the conditional
probabilities of the last category of the response variables. Doing
this, the last class of the LM model may be interpreted as that of
subjects with high tendency to use marijuana. Once the output was
post-processed, we estimated the parameters of the model by the
ergodic averages, taken over the final 800,000 iterations; the
resulting parameter estimates are reported in Table \ref{tab:est}.
We can see that these results are very similar to that obtained by
\cite{bart:06} with the EM algorithm.
\begin{table}[h!]\centering\vspace*{0.25cm}
{\small
\begin{tabular}{cccccc}
\toprule
$u$ & \centering{$\hat{\pi}_u$} & $v$ & \multicolumn{3}{c}{$\hat{\pi}_{v|u}$} \\
 & & &  \multicolumn{1}{c}{$u=1$} &  \multicolumn{1}{c}{$u=2$}&
 \multicolumn{1}{c}{$u=3$}\\
\midrule

1 & 0.868 & 1 &0.847 & 0.128 & 0.025\\
2 & 0.080 & 2 & 0.073 & 0.693 & 0.233 \\
3 & 0.052 & 3 & 0.016 & 0.065 & 0.919 \\
\bottomrule
\end{tabular}}
\caption{\em Estimated initial probabilities and transition
probabilities under the basic LM
model}\label{tab:est}\vspace*{0.25cm}
\end{table}%

We also tried to put constraints on the distribution of the response
variables, so as to give a proper interpretation of the model. We
assumed a parametrization for the conditional local logits of any
response variable given the latent state:
\begin{equation}\label{eq:par}
\eta_{y|u} = \log\frac{\phi_{y|u}}{\phi_{y-1|u}} = \zeta_u +
\omega_y, \quad u=1,\ldots,k,\; y=1,\ldots,l-1,
\end{equation}
where $\zeta_u$ may be interpreted as the tendency to use marijuana
for a subject in the state $u$ and $\omega_y$, for $y=1,\ldots,l-1$,
are the cutpoints common to all the response variables. Note that we
assumed again the constraint (\ref{eq:const1}), i.e. the
distribution of the response variables does not depend on time. The
parametrization used here requires the choice of the prior
distributions on $\zeta_u$ and $\omega_y$, that we assumed to be
$N(0,\sigma^2_\zeta)$ and $N(0,\sigma^2_\omega)$, respectively. We
considered two choices of the prior parameters, i.e. $\sigma^2_\zeta
= \sigma^2_\omega = 5, 10$, for all $u=1,\ldots,k$ and
$y=1,\ldots,l-1$, in order to see how the posterior distribution of
the number of states changes with different values of the
hyperparameters. The sensitivity to prior specification and
therefore the choice of the hyperparameters is in fact one of the
difficulties in Bayesian modeling, especially when there is little
information to be used. We computed the posterior distribution of
$k$ also considering  two different values of the parameters
$\delta_{uv}$ for the transition probabilities, i.e. $\delta_{uv} =
k\cdot I(u=v)+0.6\cdot I(u \ne v)$\footnote{indicated in Table
\ref{tab:prio} as $(k-0.6)$} and $\delta_{uv} = 1$. The prior
parameters for the initial probabilities were left unchanged, i.e.
$\delta_u = 1$ for $u=1,\ldots,k$.

In the MH step, the elements of both the parameters $\zeta_u$ and
$\omega_y$, were updated through a normal random walk proposal,
$N(0,0.5)$; moreover in the split move the parameter $\zeta_{u_0}$
was split into $\zeta_{u_1} = \zeta_{u_0} - \varphi_u$ and
$\zeta_{u_1} = \zeta_{u_0} + \varphi_u$, where $\varphi_u \sim
N(0,0.2)$, with $u=1,\ldots,k$, whereas in the reverse combine move
the selected two parameters were combined into $\zeta_{u_0} =
(\zeta_{u_1} + \zeta_{u_2})/2$. The parameters setting for the
unnormalized initial and transition probabilities was the same as in
the basic LM model. The results were based on 1,000,000 iterations
of the algorithm after a burn-in of 200,000 sweeps.

Table \ref{tab:prio} shows the results of the sensitivity analysis
to prior specification. We can see that almost all the values of the
hyperparameters lead to choose again a model with three latent
states, even if the posterior probabilities of $k$ are quite
different. The first column of the table shows our choice in this
application, that it seems to be the more adequate given the prior
information we have.
\begin{table}[h!]\centering\vspace*{0.25cm}
{\small
\begin{tabular}{rrrrrr}
\toprule
k  & \footnotesize$\delta_{uv} = (k-0.6)$ &\footnotesize $\delta_{uv} = (k-0.6)$ & \footnotesize$\delta_{uv}=1$ & \footnotesize$\delta_{uv}=1$  \\
  & $\sigma^2_\zeta = \sigma^2_\omega = 5$
  &\footnotesize $\sigma^2_\zeta = \sigma^2_\omega = 10$ &\footnotesize $\sigma^2_\zeta = \sigma^2_\omega = 5$ &\footnotesize $\sigma^2_\zeta = \sigma^2_\omega = 10$ \\
 \midrule
$ \leq 2$  & 0.000 &  0.000 &  0.000 &  0.000 \\
3   & 0.474 &  0.341 &  0.932 &  0.915    \\
4   & 0.365 &  0.361 &  0.067 &  0.082 \\
5  & 0.122 &  0.189 &  0.001 &  0.003 \\
6   & 0.031 &  0.075 &  0.000 &  0.000\\
7   & 0.007 &  0.025 &  0.000 &  0.000\\
$ \geq 8$  & 0.001 &  0.010 &  0.000 &  0.000\\
\bottomrule
\end{tabular}}
\caption{\em Posterior distribution of the number of states $k$ for
different choices of the hyperparameters under the constrained LM
model} \label{tab:prio}\vspace*{0.25cm}
\end{table}

\begin{figure}[!h]\centering\vspace*{0.25cm}
\includegraphics[scale = 0.4]{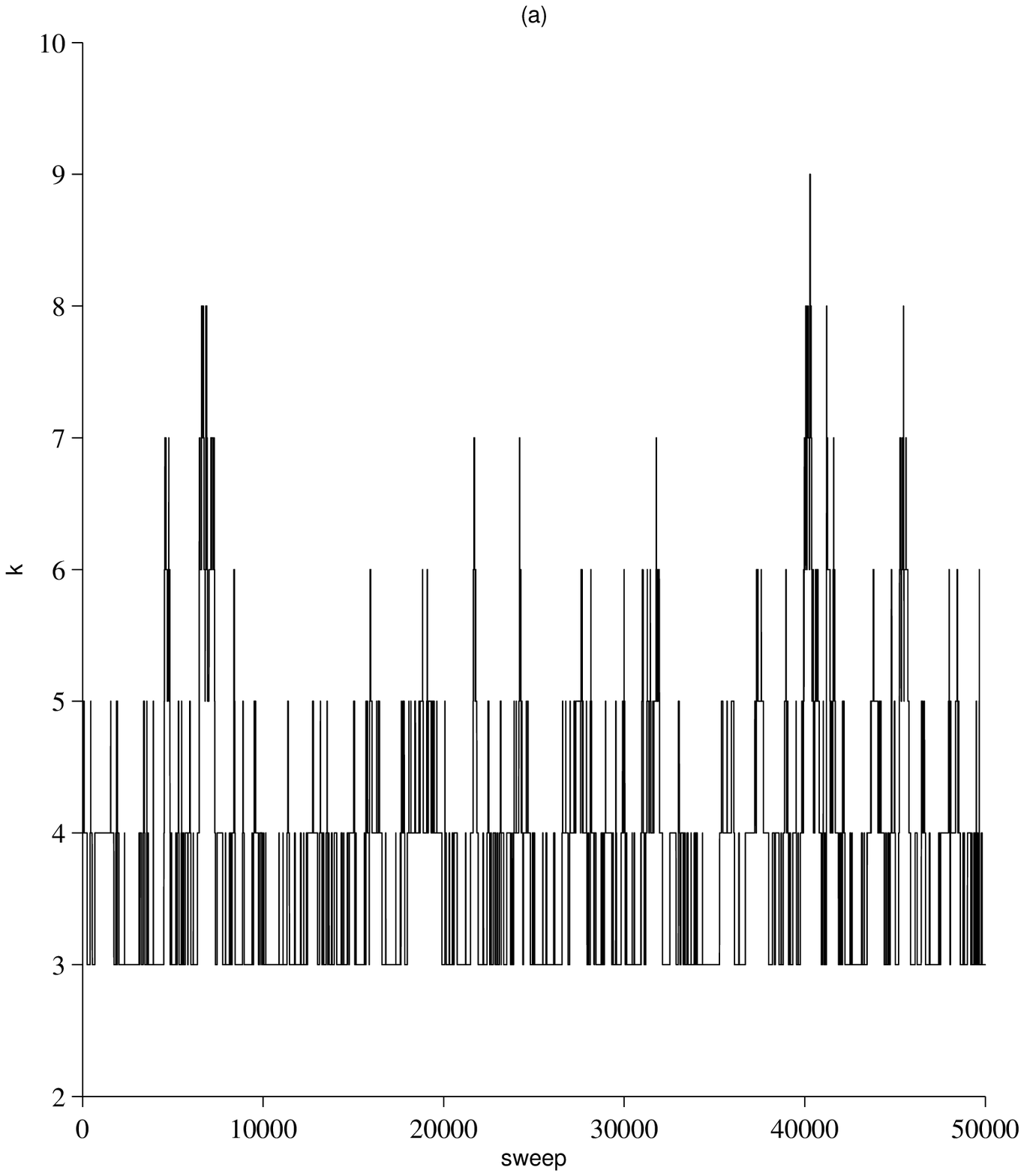}
\includegraphics[scale = 0.4]{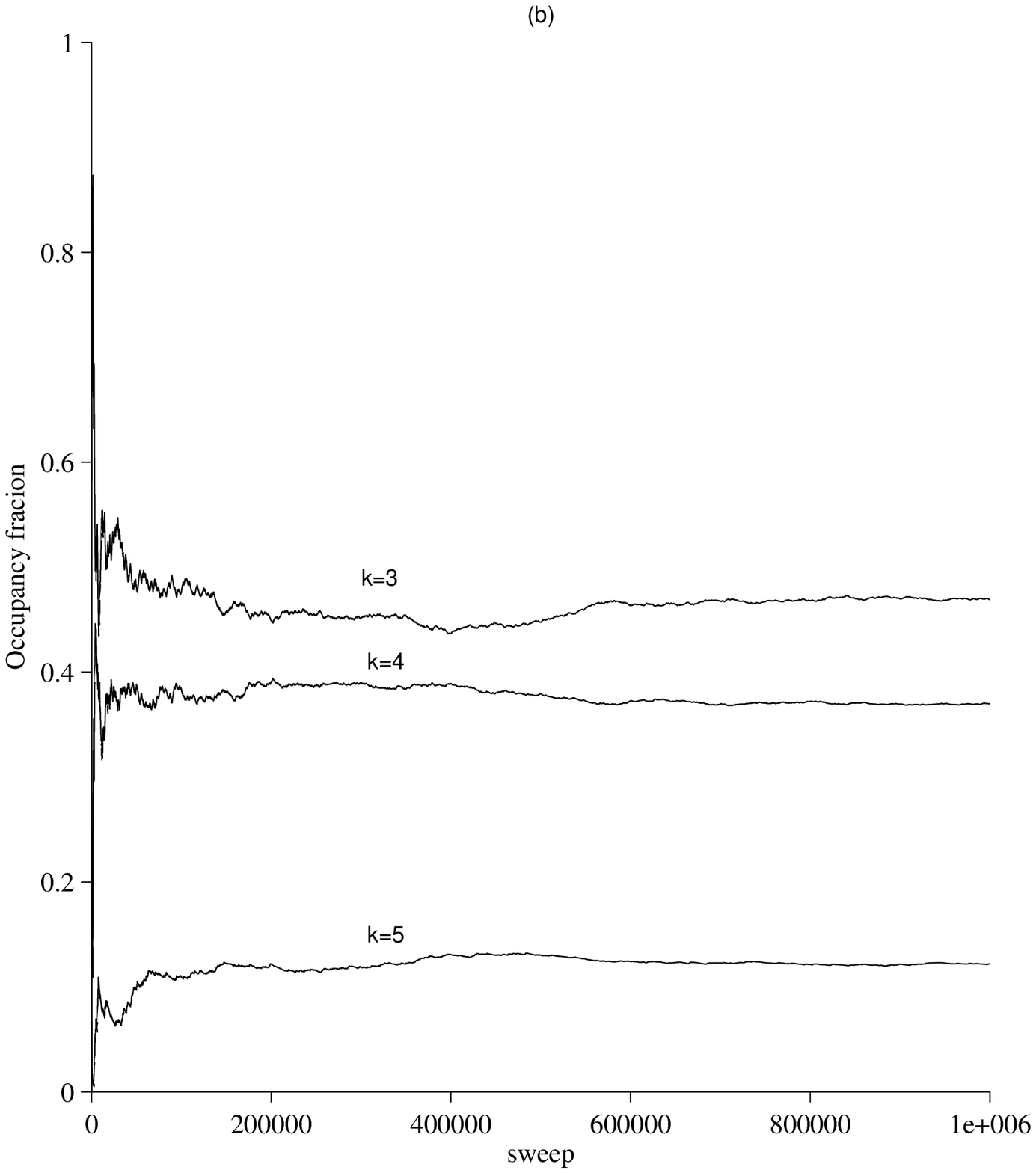}
\caption{\em(a) Number of latent states in the first 50,000
iterations after the burn-in (b) Cumulative occupancy fractions for
$k=3,4,5$ under the constrained LM model with $\sigma^2_\zeta =
\sigma^2_\omega = 5$ and $\delta_{uv} = k\cdot I(u=v)+0.6\cdot I(u
\ne v)$} \label{fig:1}\vspace*{0.25cm}
\end{figure}

The acceptance rates for the random walk MH move and for the
dimension changing moves are illustrated in Table \ref{tab:acc2}. We
can see that these rates, for split/combine and birth/death moves,
are higher than those achieved in fitting the basic LM model.
\begin{table}[h!]\centering\vspace*{0.25cm}
{\small
\begin{tabular}{lrrr}
\toprule
 &Performed &Accepted & \% Accepted\\
\midrule
MH with fixed k & 1,000,000 & \\
{\em \quad Initial probabilities} & &    195,551 & 19.56
 \\
{\em \quad Transition probabilities} & & 129,441 &12.94\\
\quad $\zeta_u$ &  & 176,501 & 17.65\\
\quad $\omega_y$ & & 185,274 & 18.53\\
Birth &  250,348 & 5,543 &  2.21 \\
Death &  249,399& 5,481 &  2.20  \\
Split &   250,199 &1,611  & 0.64\\
Combine & 250,054& 1,677  & 0.67\\
\bottomrule
\end{tabular}}
\caption{\em Acceptance rates for the MH move, the split/combine and
the birth/death move under the constrained LM model with
$\sigma^2_\zeta = \sigma^2_\omega = 5$ and $\delta_{uv} = k\cdot
I(u=v)+0.6\cdot I(u \ne v)$}\label{tab:acc2}\vspace*{0.25cm}
\end{table}
Figure \ref{fig:1} shows the mixing and the stationarity of the
algorithm with the plot of the first 50,000 values of $k$ after the
burn-in, and the plot of the cumulative occupancy fractions for
different values of $k$ against the number of sweeps. From Figure
\ref{fig:1}(b) we can see that the burn-in is adequate to achieve
stability in the occupancy fractions.

Tables \ref{tab:est1} and \ref{tab:est2} show the parameter
estimates, computed after post-processing the MCMC output. Even in
this case, the latent states may be ordered, representing subjects
with ``no tendency to use marijuana'', ``incidental users of
marijuana'' and ``high tendency to use marijuana''.
\begin{table}[h!]\centering\vspace*{0.25cm}
{\small
\begin{tabular}{rrrrrrr}
\toprule
$u$ & & $\hat{\zeta}_u$ & & $y$ & & $\hat{\omega}_y$ \\
\midrule
1 & & -5.321  & & 1 & &  0.775\\
2 & & -0.176 & & 2 & &  -1.977\\
3 & &  4.173 & & & &\\
\bottomrule
\end{tabular}}
\caption{\em Estimates of the parameters $\zeta_u$ and $\omega_y$
under the constrained LM model with $\sigma^2_\zeta =
\sigma^2_\omega = 5$ and $\delta_{uv} = k\cdot I(u=v)+0.6\cdot I(u
\ne v)$}\label{tab:est1} \vspace*{0.25cm}
\end{table}%
\begin{table}[h!]\centering\vspace*{0.25cm}
{\small
\begin{tabular}{cccccc}
\toprule
$u$ & \centering{$\hat{\pi}_u$} & $v$ & \multicolumn{3}{c}{$\hat{\pi}_{v|u}$} \\
 & & &  \multicolumn{1}{c}{$u=1$} &  \multicolumn{1}{c}{$u=2$}&
 \multicolumn{1}{c}{$u=3$}\\
\midrule
1 & 0.897  & 1 & 0.838 & 0.148  & 0.015\\
2 & 0.077 & 2 & 0.056 & 0.717  & 0.227 \\
3 & 0.026 & 3 &  0.027 & 0.058 & 0.915 \\
\bottomrule
\end{tabular}}
\caption{\em Estimated initial probabilities and transition
probabilities under the constrained LM model with $\sigma^2_\zeta =
\sigma^2_\omega = 5$ and $\delta_{uv} = k\cdot I(u=v)+0.6\cdot I(u
\ne v)$}\label{tab:est2}\vspace*{0.25cm}
\end{table}%

From the results, we can see that most subjects starts with a low
tendency to drug consumption but from the estimated marginal
probabilities of the latent classes emerge that the tendency to use
marijuana increases with age, since the probability of the third
class increases across time. From the estimated transition matrix we
can see that a large percentage of subjects remains in the same
latent class, but around 23\% of incidental users switches to the
class of high frequency users.
%
%


\subsection{Analysis of the {\em Panel Study of Income Dynamics} dataset}\label{sub:PSID}

The second dataset analyzed in this paper is very similar to that
used by \cite{hys:99} and by \cite{bart:farc:09}. The dataset was
extracted from the database derived from the Panel Study of Income
Dynamics, which is primarily sponsored by the National Science
Foundation, the National Institute of Aging, and the National
Institute of Child Health and Human Development and is conducted by
the University of Michigan. The database is freely accessible from
the website {\em http://psidonline.isr.umich.edu}, to which we refer
for details.

Our dataset concerns $n = 482$ women who were followed from 1987 to
1993. There are two binary response variables: {\em fertility}
(indicating whether a woman had given birth to a child in a certain
year) and {\em employment} (indicating whether she was employed).
The covariates are {\em race} (dummy variable equal to 1 for a black
woman), {\em age} (in 1986), {\em education} (year of schooling),
{\em child 1-2} (number of children in the family aged between 1 and
2 years, referred to the previous year), {\em child 3-5}, {\em child
6-13}, {\em child 14-} and {\em income of the husband} (in dollars,
referred to the previous year).

In analyzing the dataset, the most interesting question concerns the
direct effect of the covariates on the response variables. The
approach considered here, allows us to separate these effects from
the effect of the unobserved heterogeneity by modeling the latter by
a latent process. In this way, we admit that the unobserved
heterogeneity effect on the response variable is time-varying.

On these data, we fitted a model formulated on the basis of
assumption (\ref{eq:assump}) with $\b X_i^{(t)} = \b I_2 \otimes [\b
x_i^{(t)}]\tr$, where $\b I_d$ denoting an identity matrix of
dimension $d$ and the vector $\b x_i^{(t)}$ includes the covariates
indicated previously further to a dummy variable for each year. In
particular, the logits may be parameterized as follows:
\begin{eqnarray*}
\log\frac{\phi_{i1,1|u}^{(t)}}{\phi_{i1,0|u}^{(t)}} & = & \xi_{1u} +
[\b x_i^{(t)}]\tr \b\beta_1, \\
\log\frac{\phi_{i2,1|u}^{(t)}}{\phi_{i2,0|u}^{(t)}} & = & \xi_{2u} +
[\b x_i^{(t)}]\tr \b\beta_2,
\end{eqnarray*}
whereas, for the log-odds ratio we have
$$\small \log \frac{p(Y_{i1}^{(t)} = 1, Y_{i2}^{(t)} = 1 | U_i^{(t)} =
u,\b x_i^{(t)})}{p(Y_{i1}^{(t)} = 1, Y_{i2}^{(t)} = 0 | U_i^{(t)} =
u,\b x_i^{(t)})}+\log \frac{p(Y_{i1}^{(t)} = 0, Y_{i2}^{(t)} = 0 |
U_i^{(t)} = u,\b x_i^{(t)})}{p(Y_{i1}^{(t)} = 0, Y_{i2}^{(t)} = 1 |
U_i^{(t)} = u,\b x_i^{(t)})} = \gamma.
$$
We implemented the proposed RJ algorithm with the following
parameters setting and initialization. We used the prior
distribution defined in Section~\ref{sec:bay_basic} and
\ref{sec:bay_ext} with $\delta_u=1$, $\delta_{uv} = k\cdot
I(u=v)+0.6\cdot I(u\ne v)$ and $\sigma^2_\xi = \sigma^2_\nu = 5,10$,
with $\b\nu =(\b\be\cup \b\gamma)$. The parameters used for the
proposal distributions in the MH move were $\tau_{\la} = 0.1$,
$\tau_{\La} = 0.05$, $\tau_{\xi} = 0.5$, and $\tau_\nu  = 0.1$.
These values allowed us to obtain acceptance rates in the range
0.15-0.30. In the split/combine move we used $\tau\tr_\xi = 2$ for
the Normal proposal and $a_\la = a_\vartheta = b_\la = b_\vartheta =
1$ for the Gamma distributions. The Markov chain was initialized
from the maximum likelihood estimation obtained through the EM
algorithm, with $k = 1$. Moreover, we ran the RJ algorithm for
1,000,000 iterations discarding the first 200,000 as burn-in.

\begin{figure}[!h]\centering\vspace*{0.25cm}
\includegraphics[scale = 0.4]{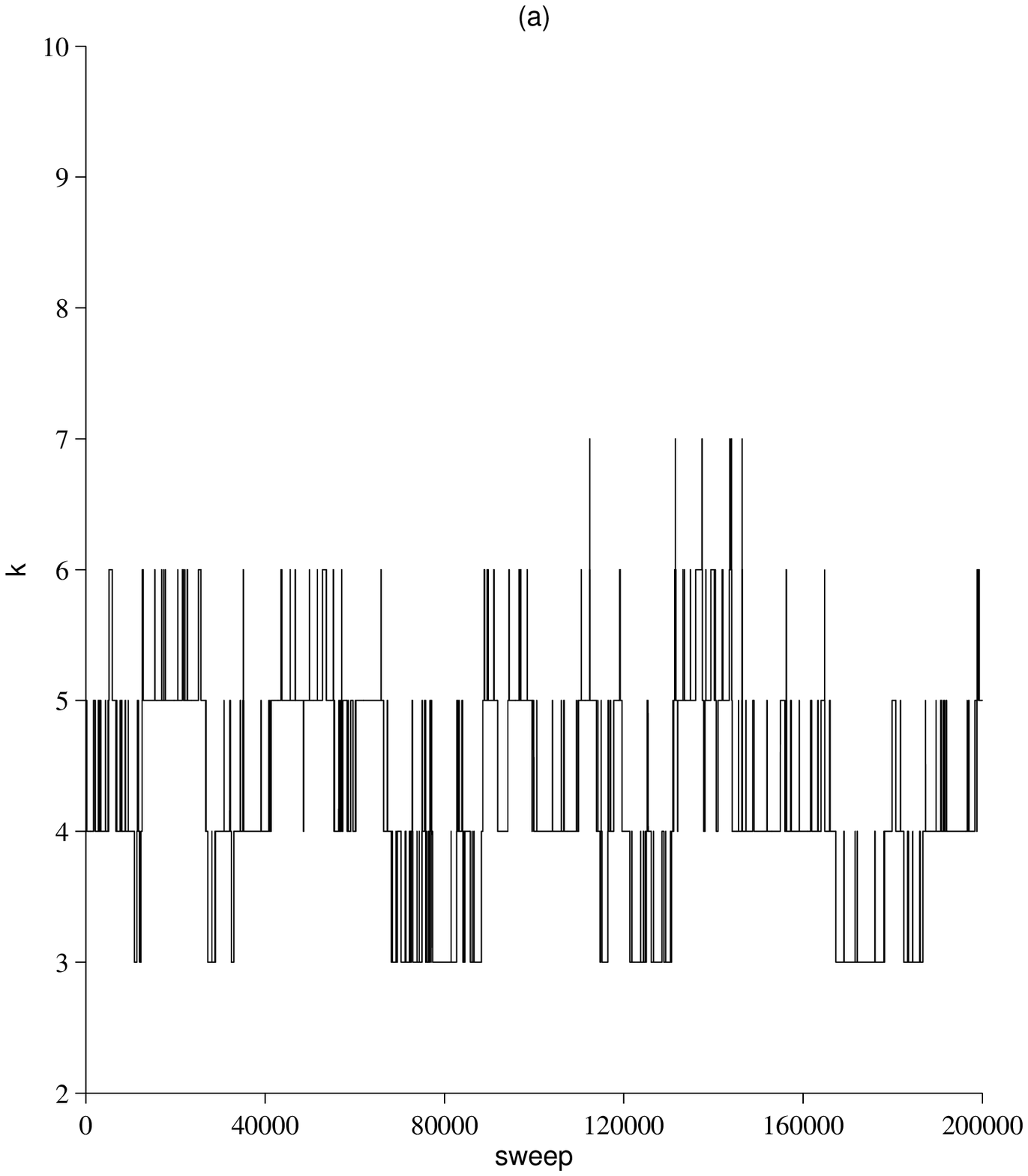}
\includegraphics[scale = 0.4]{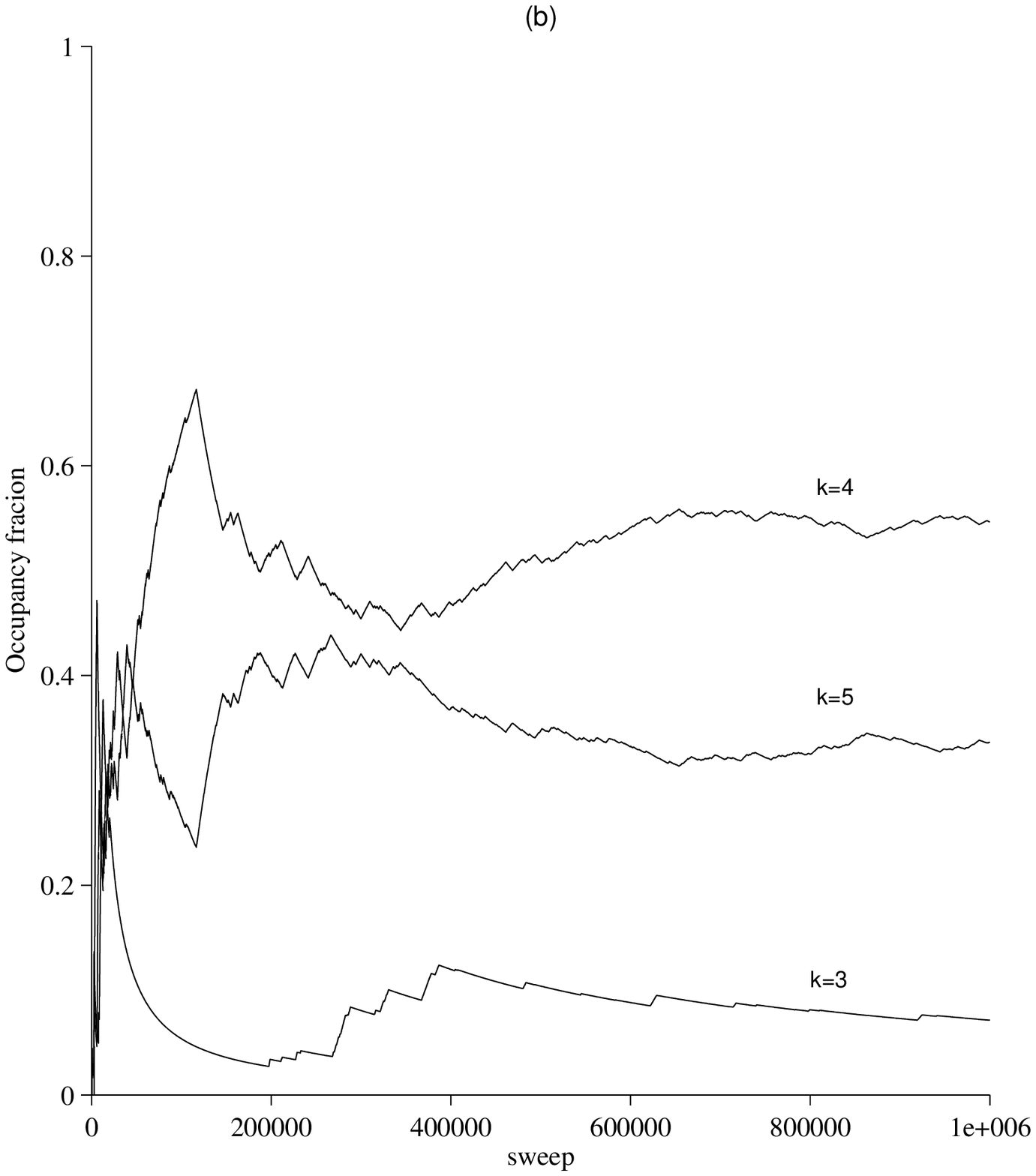}
\caption{\em(a) Number of latent states in the first 200,000
iterations after the burn-in (b) Cumulative occupancy fractions for
$k=3,4,5$ under the LM model with individual covariates for
$\sigma^2_\xi = \sigma^2_\nu = 5$}
\label{fig:mix_conv}\vspace*{0.25cm}
\end{figure}

\begin{figure}[!h]\centering\vspace*{0.25cm}
\includegraphics[scale = 0.4]{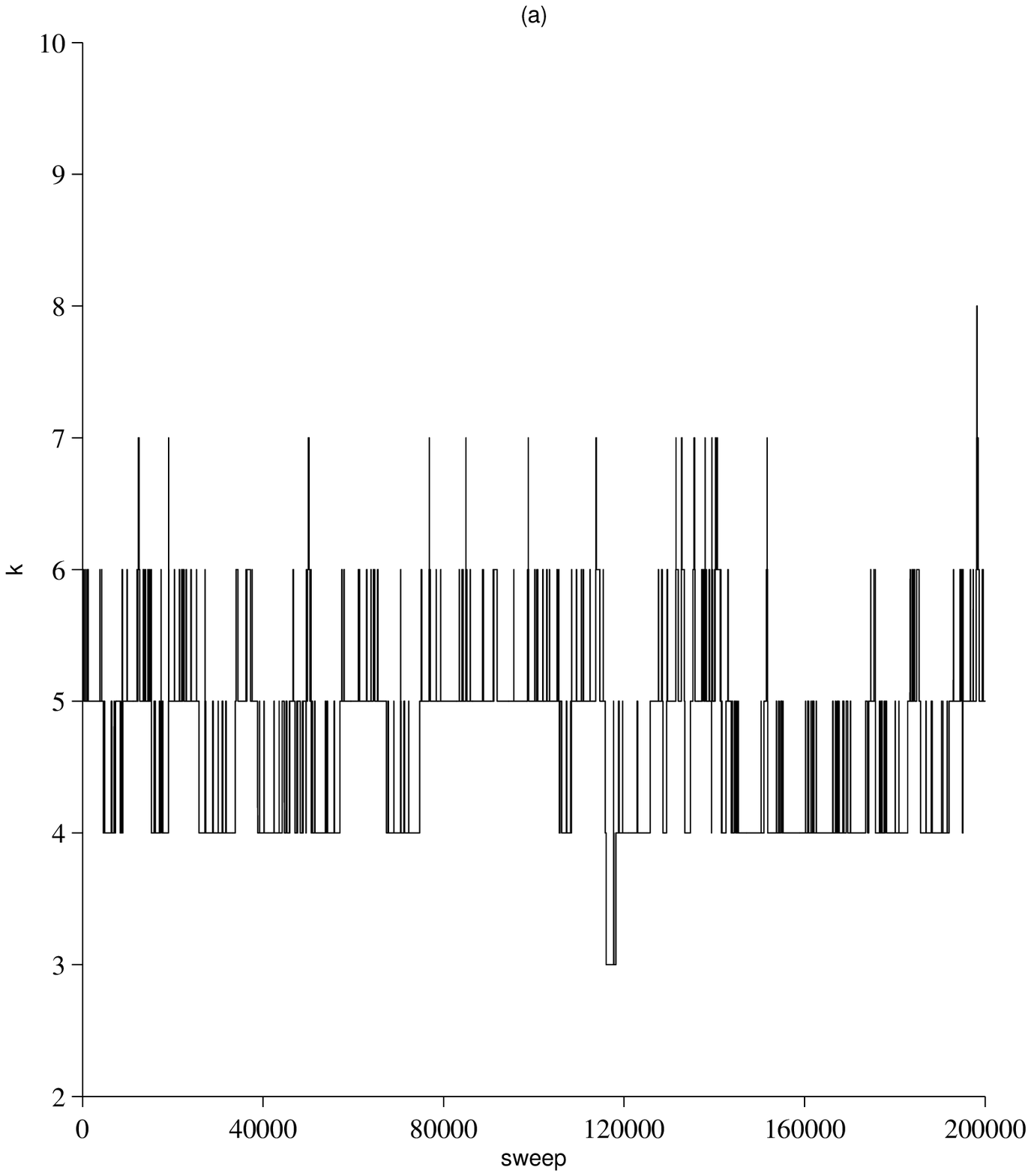}
\includegraphics[scale = 0.4]{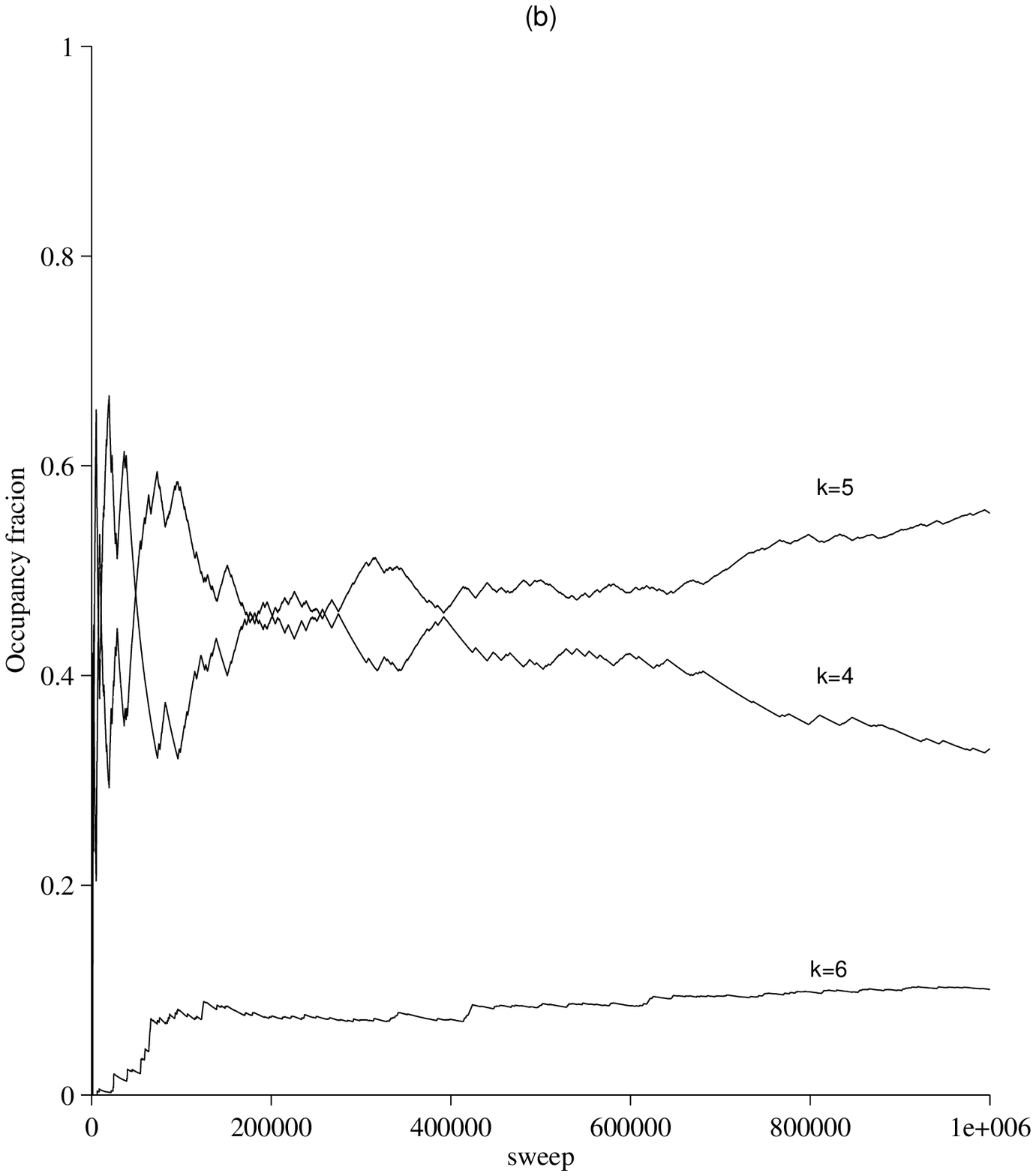}
\caption{\em(a) Number of latent states in the first 200,000
iterations after the burn-in (b) Cumulative occupancy fractions for
$k=4,5,6$ under the LM model with individual covariates for
$\sigma^2_\xi = \sigma^2_\nu = 10$}
\label{fig:mix_conv2}\vspace*{0.25cm}
\end{figure}

After the burn-in, the algorithm visited five states with posterior
probabilities illustrated in Table \ref{tab:freqv}. Table
\ref{tab:ac} also shows the acceptance rates for the different
moves, whereas in Figure \ref{fig:mix_conv} and Figure
\ref{fig:mix_conv2} are illustrated the trace of $k$ in the first
200,000 iterations, after the burn-in, and the ergodic averages of
the model probabilities, for both the choices of the prior
parameters, $\sigma^2_\xi = \sigma^2_\nu = 5,10$. We can see that
the algorithm leads to choose a model with a number of state between
4 and 5. The acceptance rates, especially for the split/combine
move, are again a bit low, but this can be due to the complexity of
the model.
\begin{table}[h!]\centering\vspace*{0.25cm}
{\small
\begin{tabular}{rrr}
\toprule k & $\sigma^2_\xi = \sigma^2_\nu = 5$ & $\sigma^2_\xi =
\sigma^2_\nu = 10$ \\
 \midrule
$\leq 3$ & 0.081 & 0.006 \\
4        & 0.554 & 0.298 \\
5        & 0.320 & 0.580  \\
6        & 0.044 & 0.107   \\
$\geq 7$ & 0.003 & 0.009  \\
\bottomrule
\end{tabular}}
\caption{\em Posterior probabilities of the number of latent states
under the LM model with individual covariates for  $\sigma^2_\xi =
\sigma^2_\nu = 5,10$.} \label{tab:freqv} \vspace*{0.25cm}
\end{table}%
\begin{table}[h!]\centering\vspace*{0.25cm}
{\small
\begin{tabular}{lrr}
\toprule
\vspace{2mm} & $\sigma^2_\xi = \sigma^2_\nu = 5$ & $\sigma^2_\xi = \sigma^2_\nu = 10$\\
 \cline{2-3}
& \% Accepted & \% Accepted \\
\midrule
MH with fixed k &   \\
{\em \quad Initial probabilities} & 19.21 & 18.64 \\
{\em \quad Transition probabilities} & 18.18 & 16.17 \\
\quad $\b\nu = (\b\beta \cup \b\gamma)$ & 16.13 & 17.71 \\
\quad $\b\xi_u$& 25.34 & 29.37\\
Birth & 0.32 &0.35 \\
Death &0.33 & 0.36 \\
Split &0.12 & 0.11\\
Combine &0.12 & 0.10\\
\bottomrule
\end{tabular}}
\caption{\em Acceptance rates for MH move, split/combine and
birth/death move under the LM model with individual covariates for
$\sigma^2_\xi = \sigma^2_\nu = 5,10$}\label{tab:ac}\vspace*{0.25cm}
\end{table}%

%
In Table~\ref{tab:par} we show the estimates of the parameters,
collected in vectors $\b\be$ and $\b\gamma$, affecting the marginal
logits of fertility and employment and the log-odds ratio between
these variables, for $\sigma^2_\xi = \sigma^2_\nu = 5$ and $k=4$ and
for $\sigma^2_\xi = \sigma^2_\nu = 10$ and $k=5$. These estimates
are straightforward to compute, through the ergodic averages of the
draws obtained when the best model was visited. In
Table~\ref{tab:par2} are also illustrated the same estimates
computed by the ergodic means over all the draws (after discarding
the burn-in) without limiting these averages to the draws obtained
when the selected model was visited. This can be done since the
parameters $\b\be$ and $\b\gamma$ do no depend by the number of
states. In both the tables, we also show the 90\%, 95\% and 99\%
posterior credible intervals not containing zero. Note that some
covariates have been standardized, before starting the estimation
algorithm, for computing purposes.
\begin{table}[h!]\centering\vspace*{0.25cm}
{\small
\begin{tabular}{llrp{0.0001mm}rp{0.0001mm}}
\toprule
 & Effect &$\sigma^2_\xi = \sigma^2_\nu = 5$ &&$\sigma^2_\xi = \sigma^2_\nu = 10$ & \\
 &        & $k=4$ && $k=5$&\\
\midrule
Logit fertility & intercept & next table && next table& \\
 & race                     & -0.082 & & -0.082&\\
 & age$^\dagger$            &  1.434 &*&  1.680&*\\
 & $\mbox{age}^{2\dagger}$  & -2.473&*** & -2.736&***\\
 & education$^\dagger$      &  0.369&*** &  0.363&*** \\
 & child 1-2                & -0.019& & -0.010& \\
 & child 3-5                & -0.375&***& -0.381&***\\
 & child 6-13               & -0.673&***& -0.683&***\\
 & child 14-                & -0.454&* & -0.467&* \\
 & income$^\dagger$         &  0.051  &&   0.057&\\
Logit employment & intercept & next table && next
table &\\
& race                     & 0.089  &&  0.205&\\
 & age$^\dagger$            &  -0.780& & -2.016& \\
 & $\mbox{age}^{2\dagger}$  & 0.367& &  1.433& \\
 & education$^\dagger$      &  1.093&***&  1.510 &*** \\
 & child 1-2                &  -0.916&*** & -1.147&*** \\
 & child 3-5                & -0.706&**&-0.894&**\\
 & child 6-13               & -0.259&  & -0.377&\\
 & child 14-                & 0.363& & 0.354& \\
 & income$^\dagger$         & -0.537&*** & -0.696&***\\
Log-odds ratio & intercept &-1.165& &-2.622&* \\
 \bottomrule
\multicolumn{6}{l}{\scriptsize $^\dagger$ In standardized form}\\
\multicolumn{6}{l}{\scriptsize $*$  posterior $90\%$HPD not
containing
zero}\\
\multicolumn{6}{l}{\scriptsize $**$  posterior $95\%$HPD not
containing zero}\\
\multicolumn{6}{l}{\scriptsize $***$  posterior $99\%$HPD not
containing zero}
\end{tabular}}
\caption{\em Posterior estimates of the model parameters affecting
the marginal logits for fertility and employment and the log-odds
ratio, under the LM model with individual covariates}
\label{tab:par}\vspace*{0.25cm}
\end{table}%
\begin{table}[h!]\centering\vspace*{0.25cm}
{\small
\begin{tabular}{llrp{0.001mm}rp{0.001mm}r}
\toprule
 & Effect &$\sigma^2_\xi = \sigma^2_\nu = 5$& &$\sigma^2_\xi = \sigma^2_\nu = 10$& &\\
 &        & $k=4$ & & $k=5$&&\\
\midrule
Logit fertility & intercept & next table& & next table& &\\
 & race                     & -0.085&& -0.080 &&\\
 & age$^\dagger$            &  1.339& & 1.725& &\\
 & $\mbox{age}^{2\dagger}$  & -2.367&**& -2.793&**&\\
 & education$^\dagger$      &  0.364&***&   0.367&*** &\\
 & child 1-2                &  -0.021& & -0.013 & &\\
 & child 3-5                & -0.378&***& -0.373&***&\\
 & child 6-13               & -0.679& ***&  -0.680 &***&\\
 & child 14-                & -0.471&*&  -0.449& &\\
 & income$^\dagger$ &   0.053& &   0.051&&\\
Logit employment & intercept & next table && next
table& &\\
& race                     & 0.079& &  0,127&&\\
 & age$^\dagger$            &  -0.849& &  -1,182&& \\
 & $\mbox{age}^{2\dagger}$  & 0.468& & 0,645&&\\
 & education$^\dagger$      &   1.145&*** &  1,395&*** &\\
 & child 1-2                & -0.870&** &  -1,151&***&\\
 & child 3-5                & -0.704&** &  -0,897&**&\\
 & child 6-13               &  -0.255&  &  -0,384&&\\
 & child 14-                & 0.352& &  0,382 & &\\
 & income$^\dagger$         &  -0.543&***  &-0,669&***&\\
Log-odds ratio & intercept &-1.594&* &-2,118 &&\\
 \bottomrule
\multicolumn{6}{l}{\scriptsize $^\dagger$ In standardized form}\\
\multicolumn{6}{l}{\scriptsize $*$  posterior $90\%$HPD not
containing
zero}\\
\multicolumn{6}{l}{\scriptsize $**$  posterior $95\%$HPD not
containing zero}\\
\multicolumn{6}{l}{\scriptsize $***$  posterior $99\%$HPD not
containing zero}
\end{tabular}}
\caption{\em Posterior estimates of the model parameters, under the
LM model with individual covariates, computed by the ergodic means
over all the iterations} \label{tab:par2}\vspace*{0.25cm}
\end{table}

On the basis of the estimates of the parameters for the covariates,
we can see that the results are very similar both if we take the
means of the draws limited to the model of interest or if we take
the overall means. Moreover these estimates are not influenced by
the prior specification we used. In particular, age seems to have an
effect on fertility but not on employment. At this regards we can
consider that the women in the sample were aged between 18 and 47,
which is a limited range of years if we want to effectively study
the effect of aging on the probability of having a job. We also note
that the education has a significant effect on both fertility and
employment, whereas income of the husband affects only the logit of
employment. Moreover, the number of children aged between 1 and 5
years has an effect on the employment while the number of children
aged between 3 and 13 years affects the fertility. The log-odds
ratio between the two response variables is negative and is
significant based on the 90\% posterior interval. This result can be
interpreted as a negative association between the two response
variables, referred to the same year.

In order to estimate the value of the remaining parameters, and in
order to face the label switching problem we applied the
post-processing algorithm of \cite{marin:et:al:05};  moreover, we
sorted the output of the algorithm on the basis of the drawn support
points $\b\xi_1,\ldots,\b\xi_k$. Once the post-processing algorithm
has been performed we could compute the estimates of those
parameters that are affected by the number of state on the basis of
the ergodic averages, after discarding the draws obtained during the
burn-in period.


In Table \ref{tab:supp} and \ref{tab:supp2} we show the results of
this estimation procedure, through the estimates of the support
points (one for the marginal logit of fertility and the other for
that of employment) corresponding to each latent state, and the
estimated initial probabilities and transition probability matrix,
for both the hyperparameters chosen in the prior specification.
Though the number of states selected is different, both the
specifications lead to the same conclusions. In particular, the
latent process can be interpreted as an error component which
follows a process that may be seen as a discrete version of an
$AR(1)$. The support points are in increasing order on the basis of
the marginal logit of employment; the latent states may therefore be
interpreted as different levels to give birth to a child or to get a
job position. For example, the first latent state corresponds to
subjects with the highest propensity to fertility and the lowest
propensity to employment. Moreover, it is interesting to observe
that the transition matrix has an almost symmetric structure, with a
large percentage of subjects that remains in the same latent state.
\begin{table}[h!]\centering\vspace*{0.25cm}
{\small
\begin{tabular}{l|rr|r|rrrrr}
\toprule
& \multicolumn{2}{c|}{Support points} &    \\
Latent state  &Fertility & Empl. &\multicolumn{1}{c|}{Initial prob.} & \multicolumn{4}{c}{Transition probabilities}\\
\midrule
1 &  -1.796 & -4.937  &  0.092 &  0.734  & 0.065 & 0.055& 0.146 \\
2 &  -1.936 & -3.718  &  0.102 &  0.072  & 0.643  & 0.067& 0.219\\
3 &  -2.648 & -0.002  &  0.228 &  0.071  & 0.072  & 0.754& 0.103\\
4 &  -2.609 &  5.980  &  0.578 &  0.021  & 0.027  & 0.009&   0.944\\
\bottomrule
\end{tabular}}
\caption{\em Estimated support point for each latent state,
estimated initial probabilities and estimated transition probability
matrix for the LM model with individual covariates with
$\sigma^2_\xi = \sigma^2_\beta=\sigma^2_\gamma = 5$ and $k=4$}
\label{tab:supp} \vspace*{0.25cm}
\end{table}
\begin{table}[h!]\centering\vspace*{0.25cm}
{\small
\begin{tabular}{l|rr|r|rrrrr}
\toprule
Latent & \multicolumn{2}{c|}{Support points} &  \multicolumn{1}{c|}{Initial}   \\
 state  &Fertility  & Empl. &\multicolumn{1}{c|}{prob.} & \multicolumn{5}{c}{Transition probabilities }\\
\midrule
1  & -1.902 & -5.461 & 0.084 &  0.553 &  0.063  & 0.090 &0.060 & 0.234  \\
2  & -1.965 & -5.117 & 0.103 &  0.043 &  0.754  & 0.050 &0.052 & 0.101 \\
3  & -3.180 &  0.174 & 0.194 &  0.048 &  0.052  & 0.684 &0.163 & 0.053 \\
4  & -2.117 &  3.542 & 0.090 &  0.187 &  0.081  & 0.059 &0.555 & 0.117\\
5  & -2.634 &  7.786 & 0.530 &  0.022 &  0.014  & 0.005 &0.021 & 0.939 \\
\bottomrule
\end{tabular}}
\caption{\em Estimated support point for each latent state,
estimated initial probabilities and estimated transition probability
matrix for the LM model with individual covariates with
$\sigma^2_\xi = \sigma^2_\beta=\sigma^2_\gamma = 10$ and $k=5$}
\label{tab:supp2}\vspace*{0.25cm}
\end{table}
\section{Conclusion}\label{sec:concl}

In this paper, we proposed a framework for Bayesian inference on a
class of LM models for categorical longitudinal data. We considered
in particular the basic LM version, in which the latent Markov chain
is of first-order and time homogeneous, and some extended versions
which include constraints and individual covariates in the
measurement model, which corresponds to the conditional distribution
of the response variables given the latent states.

The proposed inferential approach is based on a system of priors
whose specification follows that adopted by \cite{capp:et:al:05} and
\cite{spez:10} for HM models. In particular, this system of priors
is formulated on a transformation of the initial and transition
probabilities which is equivalent to a system based on Dirichlet
distributions.

With the aim of estimating the model parameters and the number of
latent states, we implemented an RJ algorithm that allows us to
simultaneously draw samples from the posterior distribution of the
parameters and the number of states. The choice of the system of
priors leads to an algorithm easier to implement; in particular, the
computation of the Jacobian of the transformation from the current
value of the parameters to the new value is easier with respect to a
system of priors based on Dirchlet distributions. The structure of
the proposed RJ algorithm has many points in common with the RJ
algorithms for mixture models of \cite{rich_green:97} and for HM
models of \cite{robe:ryde:titt:00}. In particular, our algorithm is
based on moves of MH type, which update the parameters of the
current model given the number of states, and moves of split/combine
and birth/death type, aimed at also updating the number of states.


The proposed approach can be extended in several ways and even
applied to different LM formulations. We are referring, in
particular, to the development of a similar Bayesian framework for
the extended versions of the LM model in which individual covariates
are included in the latent model. These covariates are then assumed
to affect the initial and transition probabilities of the latent
Markov chain. This extension would require a different formulation
of the priors, based, for instance, on Normal distributions assumed
on suitable transformations of these initial and transition
probabilities. Natural transformations are based on multinomial
logits.

Moreover, it is possible to extend the proposed framework in order
to deal with {\em missing responses}, that we can assume to be
missing at random in the sense of \cite{rubin:76}. Thus, the missing
data mechanism is ignorable for posterior inference. It is possible
to make the proposed RJ algorithm able to handle data with missing
responses of this type. The missing data can be estimated along with
the parameters of the LM model, through the steps of the algorithm.

Other interesting extensions concern the implementation of an
algorithm for {\em path prediction}, i.e. to predict the sequence of
latent states of a subject on the basis of the observed data, and
the {\em Bayesian model averaging}, in order to estimate parameters
with invariable dimension with respect to the number of states
(e.g., parameters for the covariates) and for prediction of the
responses.

Finally, an aspect that has to be remarked and that requires
additional future work, concerns the sensibility of the inferential
results on the prior specification. A first analysis has already
been done in the two illustrative examples, showing that differences
in the prior specification may lead to differences in the estimation
of the number of latent states. However, further research is necessary in
order to have a more conclusive answer about this issue. 


\end{document}